\documentclass[useAMS,usenatbib]{mn2e}

\usepackage{graphicx}
\usepackage{natbib}

\title[A giant radio halo in the cool core cluster CL1821+643]{A giant radio halo in the cool core cluster CL1821+643}
\author[A. Bonafede et al.]{A. Bonafede,$^1$\thanks{E-mail:
annalisa.bonafede@hs.uni-hamburg.de}  H.T. Intema$^2$, M. Br\"uggen$^1$,  H. R.  Russell$^4$,  G. Ogrean$^1$, K. Basu$^3$,  
 \newauthor  M. Sommer$^3$, R.J. van Weeren$^5$, R. Cassano$^6$, A. C. Fabian$^4$, and H. J. A. R{\"o}ttgering$^7$.\\
$^1$ Hamburger Sternwarte, Universit\"at Hamburg, Gojenbergsweg 112, 21029, Hamburg, Germany. \\
$^2$ National Radio Astronomy Observatory, 1003 Lopezville Road, Socorro, NM 87801-0387, USA.\\
$^3$ Argelander Institut f\"ur Astronomie, Universit\"at Bonn, D-53121 Bonn, Germany.\\
$^4$Institute of Astronomy, Madingley Road, Cambridge CB3 0HA.\\
$^5$ Harvard-Smithsonian Center for Astrophysics, 60 Garden Street, Cambridge, MA 02138, USA.\\
$^6$ Istituto di Radioastronomia INAF, via P. Gobetti 101, 40129 Bologna, Italy.\\
$^7$ Leiden Observatory, Leiden University, 2300 RA Leiden, the Netherlands.\\
}

\begin{document}

\date{Accepted É Received...}

\pagerange{\pageref{firstpage}--\pageref{lastpage}} \pubyear{2002}

\maketitle

\begin{abstract}
 Giant radio halos are Mpc-size sources found in some merging galaxy clusters. The synchrotron emitting electrons are thought to be (re)accelerated by plasma turbulence induced by 
the merging of two massive clusters. Cool core galaxy clusters have a low temperature core, likely an indication that a major merger has not recently occurred. 
CL1821+643 is one of the strongest cool core clusters known so far. Surprisingly, we detect a giant radio halo with a largest linear size of $\sim$ 1.1 Mpc. 
We discuss the radio and X-ray properties of the cluster in  the framework of the proposed models for giant radio halos. If a merger is causing the radio emission, despite the presence of a cool-core, we suggest that it should be off-axis, or in an early phase,  or a minor one.
\end{abstract}


\section{Introduction}
Giant radio halos are extended synchrotron sources found in a fraction of massive galaxy clusters. Their surface brightness is faint ($\sim$ 1$\mu$Jy/arcsec$^2$ at 1.4 GHz) and characterised by a steep spectrum\footnote{The radio spectrum is defined as $S(\nu) \propto \nu^{-\alpha}$.} with a spectral index $\alpha>1$ (see \citealt{Feretti12} for a recent review). Radio emission on Mpc-scales requires that the emitting electrons are either (re)accelerated in situ, e.g. through merger-generated turbulence, or continuously injected into the intra-cluster medium (ICM), for instance by inelastic  hadronic collisions between relativistic and thermal protons \citep[see review by][and ref. therein]{BJ14}. Hadronic models are currently disfavoured as they fail to reproduce the steepest  spectrum radio halos, and predict $\gamma$-ray emission that has not been observed by the {\it Fermi} Satellite \citep[e.g.][]{2012MNRAS.426..956B}. At the same time, recent LOFAR observations  of Abell 2256 indicate that  the formation process of halos might be more complex than previously thought \citep{vanweeren12}.\\
\indent Mini halos are another class of radio sources that are located in cool core galaxy clusters. Their emission is limited to smaller spatial scales ($\sim$50- 500 kpc), and has a steep spectrum \citep[see][and ref. therein.]{Feretti12}. 
 Again, the radio emission could be caused by turbulence induced by gas-sloshing in the central potential well or by  inelastic collisions between relativistic and thermal protons \citep[e.g.][and references therein]{BJ14}.\\
\indent The picture that emerges from X-ray observations is that mergers between clusters are able to destroy the cool core and leave clear signatures in the emission of the gas \citep[e.g.][]{Rossetti10}.
Hence, mergers would be responsible for the cool core non-cool core dichotomy, supporting an evolutionary scenario between cool core and non-cool core clusters \citep{Rossetti11}. However, simulations show that disrupting the cool core during cluster mergers is not obvious  \citep[see e.g.][]{Poole08}. \\
\indent Since their first discovery, giant radio halos have always been found in merging galaxy clusters, characterised by the absence of a cool core, while mini halos are found in clusters hosting a cool core \citep[][and ref. therein]{Feretti12}.\\
\indent The cluster CL1821+643 surrounds the quasar H1821+643. CL1821+643  was discovered by \citet{1992AJ....103.1047S} through optical observations.
The redshift information, available for six member galaxies plus the quasar, allowed them to estimate the cluster redshift: z=0.299.
 CL1821+643 was also detected by the {\it Planck} satellite (PSZ1\,G094.00+27.41).  Through the
thermal Sunyaev Zel'dovich  (SZ) effect, \citet{Planck13} infer a mass $M_{500}=6.311 \times 10^{14} M_{\odot}$.
\citet{2010MNRAS.402.1561R}  observed the cluster with {\it Chandra}. Thanks to an accurate modelling of the quasar emission, they were able to 
separate the cluster and quasar emission, and analysed both the ICM properties down to the inner regions, and the interaction of the cluster with the powerful quasar. 
They found that the temperature of the cluster decreases from $\sim$ 9 keV  to 1.3 keV in the centre, and derived a a short radiative cooling time of $\sim$ 1 Gyr, typical of strong cool core systems.   
In the inner 100 kpc, the X-ray morphology shows extended spurs of emission from the core, a small radio cavity, and 
a weak shock or cold front at 15$''$ from the cluster centre. 
The optical quasar hosts a  FRI radio source, 
which is not clearly related to the X-ray gas morphology \citep{2010MNRAS.402.1561R}.\\
\indent \citet{Wold02} have analysed the weak lensing properties of the cluster, finding that the cluster is slightly elongated towards North-West, as also suggested by the X-ray emission on large scales \citep{2010MNRAS.402.1561R}.\\
\indent In this work, we present new radio observations of CL1821+643, where we discover diffuse radio emission on Mpc scales.
In Sec. \ref{sec:observations}, we present Giant Meterwave Radio Telescope (GMRT) and archival Very Large Array (VLA) observations of the cluster. Results are presented in Sec. \ref{sec:results}, where we also analyse the dynamical status of the cluster. We discuss the results and conclude  in Sec. \ref{sec:discussion}. Throughout this Letter, we assume a $\Lambda$CDM cosmology ($H_{0}=71 \, \rm{km \, s^{-1} Mpc^{-1}}$, $\Omega_m=0.27$, $\Omega_{\Lambda}=0.73$). At the cluster redshift (z=0.299) one arc minute corresponds to 265 kpc.

\begin{table*}
\label{tab:obs}
 \centering
   \caption{Radio observations of CL1821+643.}
  \begin{tabular}{c c c c c c c c c c c }
  \hline
Pointing                            &  Obs. date    &  $\nu$    & $\Delta \nu$ & t     &$\theta_{\rm FR}$   & $\sigma_{\rm FR}$  &   $\theta_{\rm HR}$  &  $\sigma_{\rm HR}$ & $\theta_{\rm LR}$& $\sigma_{\rm LR}$  \\
RA \& DEC J2000            &                       & MHz             &  MHz    &   h      &  $'' \times ''$      & mJy/beam        &  $'' \times ''$        & mJy/beam      & $'' \times ''$  & mJy/beam \\         
18 21 57.3  +64 20 36     &  27-01-2013       &  323          &   33      &  6         &     12 $\times$9  & 0.07    & 11 $\times$7 & 0.13 & 30$\times$26   & 0.20     \\
18 21 57.2  +64 20 36      &   22-09-1996    & 1665 &         50           &  0.37    &      49 $\times$32  &   0.10    &45 $\times$30 & 0.16  &53$\times$38 &0.20\\
\hline 
\multicolumn{11}{l}{\scriptsize  Col. 1: Right ascension and declination of the pointing centre if the observation; Col. 2: Date of observation;}\\
 \multicolumn{11}{l}{\scriptsize Col. 3:  Central frequency ; Col. 4: Bandwidth; Col 5: Net on-source time;  Col 6: Restoring beam of the full-resolution image;}\\
  \multicolumn{11}{l}{\scriptsize  Col 7: rms noise of the full resolution image. Col. 8: Restoring beam of the high-resolution image; Col 9: rms noise of the high-resolution image. }\\
    \multicolumn{11}{l}{\scriptsize  Col 10:  Restoring beam of the low-resolution image. Col. 11: rms noise of the low-resolution image}
\end{tabular}
\end{table*}

\begin{table}

 \centering
   \caption{Properties of the giant radio halo.}
  \begin{tabular}{c c c c }
  \hline    
Frequency    & Flux density  & LAS  & LLS \\
        MHz          &  mJy    &   $''$  & kpc \\
    323          &  62$\pm$4 & 250 &  1100 \\           1665         & 11.9$\pm$0.5 & 210 &  930  \\
\hline
 \end{tabular}
 \label{tab:sources}
\end{table}

\begin{figure}
\vspace{90pt}
\begin{picture}(90,90)
\put(-22,-15){\includegraphics[width=9cm]{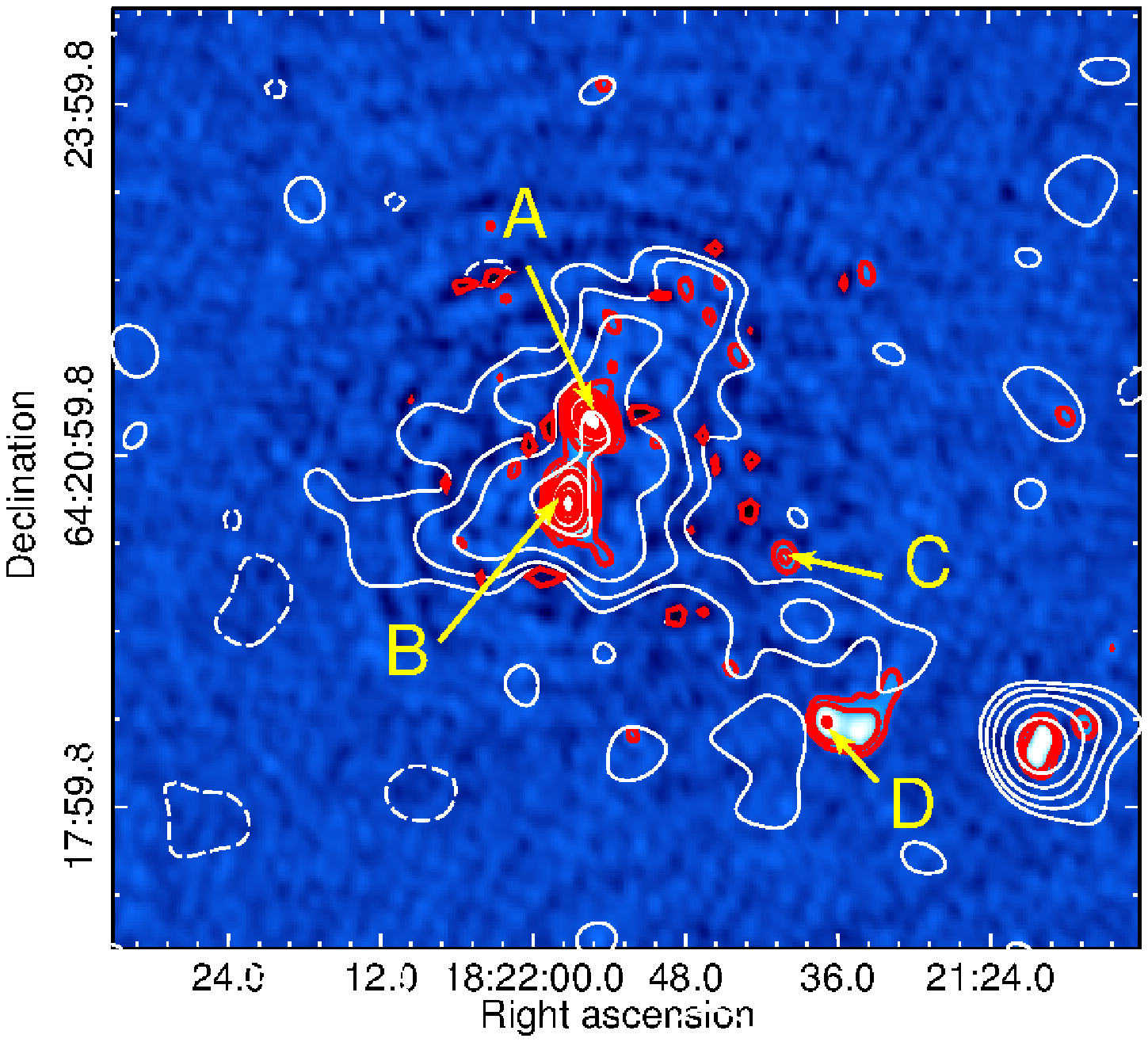}}
\put(140,87){\includegraphics[width=3.5cm]{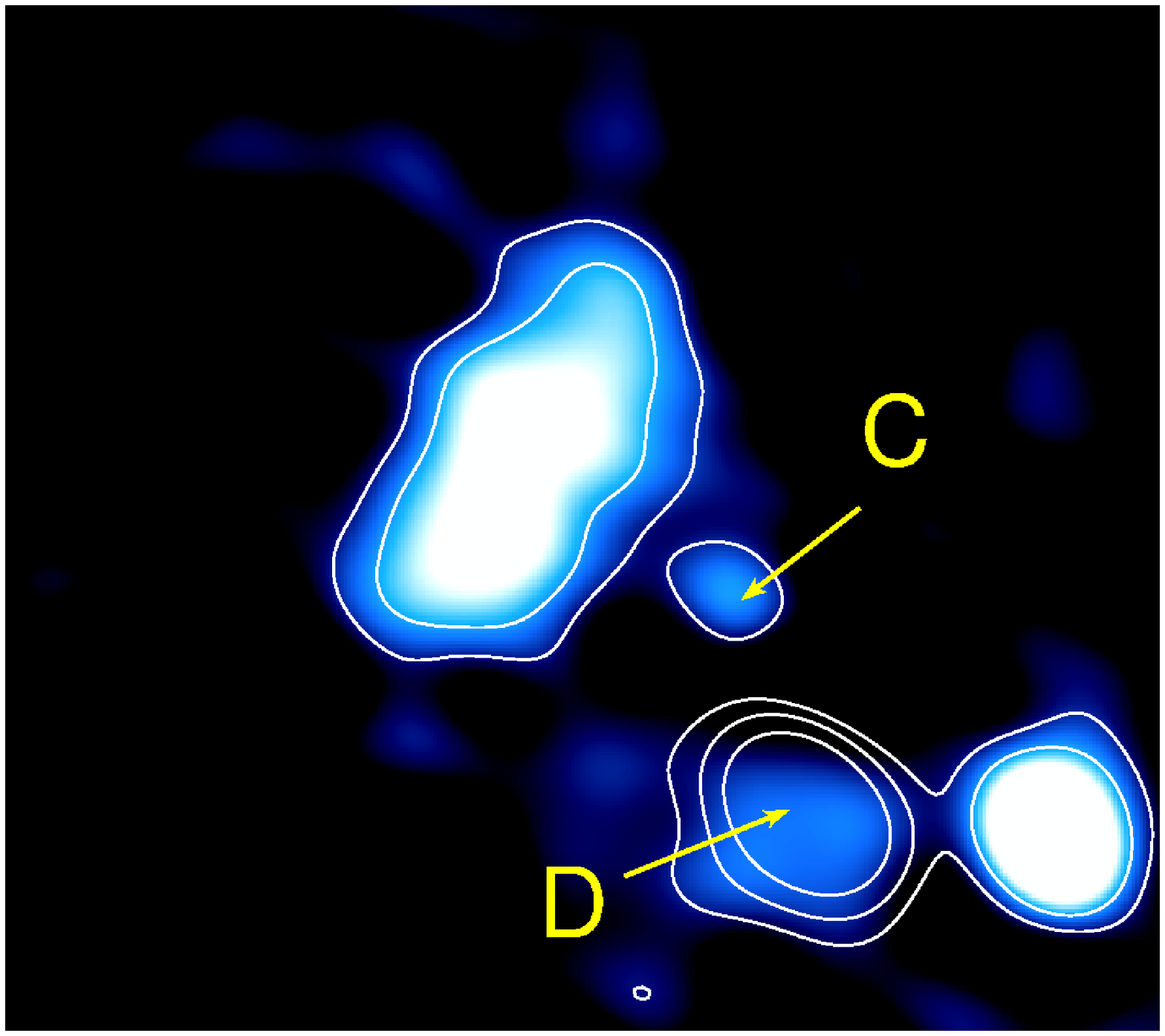}}
\end{picture}
\caption{The cluster CL1821+643. Main panel colors and red contours: HR image at 323 MHz. White contours: LR image at 323 MHz. The first contours are drawn at $3 \sigma_{\rm LR,HR}$,  other contours are spaced by a factor 2.  The $-3 \sigma_{\rm LR,HR}$ contours are dashed. In the inset: LR image at 1665 MHz in colours and contours (at 3,6,12 $\sigma_{\rm LR}$, the -3$\sigma_{\rm LR}$ contour is dashed). The inset shows the same region as the main panel (A and B have been subtracted).}
\label{fig:HR}
\end{figure} 

\section{Radio Observations}
\label{sec:observations}
We have observed the cluster CL1821+643 with the GMRT in the framework of a larger project aimed at discovering diffuse emission in clusters detected by {\it Planck}.
Observations were carried out at 323 MHz, using a 33 MHz bandwidth subdivided into 256 channels and
8s integration time.
The sources 3C48 and 3C286 were observed for 15 minutes at the beginning of the observing block, 
and used as absolute flux and  bandpass calibrators, adopting the  \citet{ScaifeHeald12} absolute  flux scale.
The absolute flux calibrators were also used to estimate  the instrumental contribution to the
antenna gains, which is also needed for ionospheric calibration, and the instrumental phase information was used to correct the target field.
The main steps of the calibration procedure are outlined below, and are based on AIPS (NRAO Astronomical Image Processing System), SPAM \citep{SPAM} and Obit \citep{OBIT} tools. 
Strong Radio Frequency Interferences (RFIs) were removed from the target field data by statistical outlier flagging tools. Remaining low-level RFIs
were modelled and subtracted from the data using Obit.
 The dataset has been averaged down to  24 channels, a compromise to speed up the imaging process and avoid significant bandwidth smearing. For the phase calibration, we started from a model derived from the northern VLA Sky Survey (NVSS, \citealt{NVSS})
 and then proceeded with self-calibration loops. To compensate for the non-coplanarity of the array, we used the polyhedron (facet-based) wide-field imaging technique as available in AIPS. 
 We performed several rounds of imaging and self-calibration, inspecting the residual visibilities for a more accurate removal of low-level RFIs. 
To correct for ionospheric effects,  we applied SPAM calibration and imaging. 
The presence of strong sources in the field of view enables one to derive directional-dependent gains for each of them and to 
use these gains to fit a time variable phase-screen over the entire array. 
After ionospheric corrections, data, data has been imported to CASA 
(Common Astronomy Software Applications package) for further imaging and self-calibration steps. \\
\indent In addition, we have downloaded an observation performed in the L band in the D array configuration from the VLA data archive.
Using AIPS, we followed a standard procedure for the data reduction.
The source 3C286 has been used to set the  absolute flux,  using the \citet{PerleyTaylor} flux scale. As no phase-calibrator was 
scheduled during the observation, we transferred the gains from 3C286 to the target and proceeded with cycles of self-calibration
on the target field.  Details about the observation are listed in Table \ref{tab:obs}.\\
\indent For both GMRT and VLA observations, we have produced three different images of the target field:\\
$\bullet$ {Full-resolution image (FR): using all the baselines and  Briggs weighting scheme with robust $R=0$.  }\\
$\bullet$ {High-resolution image (HR): using all the baselines longer than 0.9 $k\lambda$, to filter out the emission on scales larger than 1 Mpc. Sources have been cleaned down to 1$\sigma$, $\sigma$ being the rms noise level of the image. On these images, we have identified the sources within a radius of 1 Mpc from the cluster centre (see Fig. 1). The quasar and the head-tail radio galaxy detected by \citet{2001ApJ...562L...5B} are labelled as B and A, respectively.}\\
$\bullet$ {Low-resolution image (LR): First, we subtracted from the uv-data the visibilities corresponding to the the clean components of the HR image for the sources A, B, C, and D (see Fig. \ref{fig:HR}). Then, we  imaged the uv-subtracted dataset including all baselines and tapering down the baselines longer than 4k$\lambda$ to gain sensitivity towards the diffuse emission.}\\
The final images were  corrected for the primary beam response. Furthermore, GMRT images have been corrected for the system temperature using the \citet{Haslam} all-sky map. 
We estimate that the residual amplitude errors are of the order of 6\% at 323 MHz.  VLA observations amplitude errors are $\sim$4\% at 1.665 GHz.

\section{Results}
\label{sec:results}
\subsection{Radio emission and cluster properties}
\begin{figure}
\vspace{90pt}
\begin{picture}(90,90)
\put(0,-10){\includegraphics[width=\columnwidth]{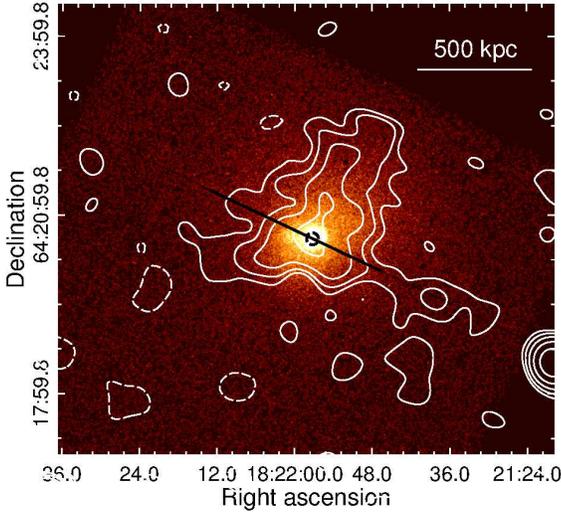}}
\end{picture}
\caption{Colors: {\it Chandra} exposure-corrected  image in the 0.5 - 7 keV energy band smoothed with a 2D Gaussian of $\sigma=$1$''$.  The readout-strip is blanked.  Contours are the same as the 323 MHz LR image in Fig. 1. The black dashed circle is centred on the quasar and has a radius of 0.05$\times R_{ap}$ (see text for details).}
\label{fig:Xradio}
\end{figure}

As shown in Fig. \ref{fig:Xradio}, diffuse radio emission is clearly detected in CL1821+643. 
We note that other compact sources, that we subtracted from the uv-data, do not leave residuals in the LR images.
To analyse its radio morphology we rely on the 323 MHz observation, which has a longer exposure time and a
wider range of baselines with respect to the 1.665 GHz one. 
In Fig. \ref{fig:Xradio} the LR image at 323 MHz is shown.
The radio emission is located at the cluster centre, and extends for $\sim$1 Mpc towards NW. 
Because of its size and location, we classify this emission as a giant radio halo.\\
Further emission is tentatively detected at $3\sigma$ significance at the west of the cluster (see Fig. \ref{fig:Xradio}).
Although it does not overlap with sources C and D (Fig. \ref{fig:HR}), deeper observations would be needed to confirm the detection.
We have measured the flux density of the giant radio halo on the LR images at 323 MHz and at 1.665 GHz. 
The error on the giant radio halo flux has been estimated as $\Delta S=\sqrt{(F_{h} Err_{cal})^2 + (\sigma_{\rm LR} \sqrt{N_{b}})^2}$, where $F_h$ is the flux density of the giant radio halo measured from the LR image, $Err_{cal}$ is the calibration error, $\sigma_{\rm LR}$ is the rms noise of the LR image, and $N_{b}$ is the number of independent beams sampling the diffuse emission. Details about the giant radio halo flux density and size are listed in Table \ref{tab:sources}.\\
\indent To estimate the spectral index of the halo, we imaged the uv-subtracted data, selecting the range of baselines that are densely sampled by both observations
(from 0.3k$\lambda$ to 5.5k$\lambda$) and convolved the images with the same beam of $\sim$ 60$'' \times 50''$. Because of the poorer sensitivity of
the 1.665 GHz observation, only the central region is detected. It has a spectral index of $\alpha = 0.87\pm 0.04$. Using these images, a lower limit for the spectral index of the entire halo can be derived, computing the flux densities in both images wherever the 323 MHz image has a signal above 3 times the noise level, obtaining  $\alpha \geq1.04 $. 
Using the fluxes measured in the LR images at 323 MHz and 1.665 GHz (Table \ref{tab:sources}), we obtain an upper limit for the spectral index: $\alpha \leq 1.1$.\\
\indent The power of giant radio halos is known to correlate with the cluster X-ray luminosity: $P_{\rm 1.4 \,GHz}-L_{\rm [0.1-2.4] \, keV}$ correlation, and with the cluster SZ signal: $\rm{SZ} - P_{\rm 1.4 \,GHz}$ correlation (\citealt{Basu12}, \citealt{cassano13}).
We have computed the power of the giant radio halo at 1.4 GHz extrapolating the flux density observed in the 323 MHz image and assuming a spectral index in the range  $1.04 \le\alpha \le1.1$. The estimated power at 1.4 GHz, $P_{\rm 1.4 \,GHz}$, is in the range  (3.6 - 3.8) $\times 10^{24}$ W Hz$^{-1}$.
We have computed $L_{\rm [0.1-2.4] \, keV}$ from the {\it Chandra} observation of the cluster presented in \citet{2010MNRAS.402.1561R}, using an absorbed MEKAL model
in the X-ray fitting package {\it xspec} 
with a temperature $\rm{T}=7.0 \pm 0.2$ keV, and a metallicity $\rm{Z}=(0.29 \pm 0.03) \, Z_{\odot}$.
We obtain $L_{\rm [0.1-2.4 \, keV]}=(1.44 \pm 0.01)\times 10^{45}$ erg s$^{-1}$, which would put the halo a factor $\sim$ 3 below the $P_{\rm 1.4 \,GHz}- L_{\rm [0.1-2.4] \, keV}$ correlation, i.e. the cluster is under luminous in radio for its X-ray luminosity. $L_{\rm [0.1-2.4 \, keV]}$ is here computed within an aperture of 650 kpc, the maximum allowed by this {\it Chandra} observation, while \citet{cassano13} computed $L_{\rm [0.1-2.4 \, keV]}$ within $r_{500}$, corresponding to $\sim$ 1.2 Mpc for this cluster. Hence, the cluster could be more than a factor 3 below the correlation. The SZ flux is a more robust estimate of the cluster mass, since it is less biased by the dynamical state of the cluster. Indeed, the cluster follows  the SZ-$P_{\rm 1.4 \,GHz}$ correlation within the errors. 

\subsection{Characterisation of the dynamical state}
\begin{figure}
\vspace{230pt}
\begin{picture}(120,120)
\put(0,-10){\includegraphics[width=8cm]{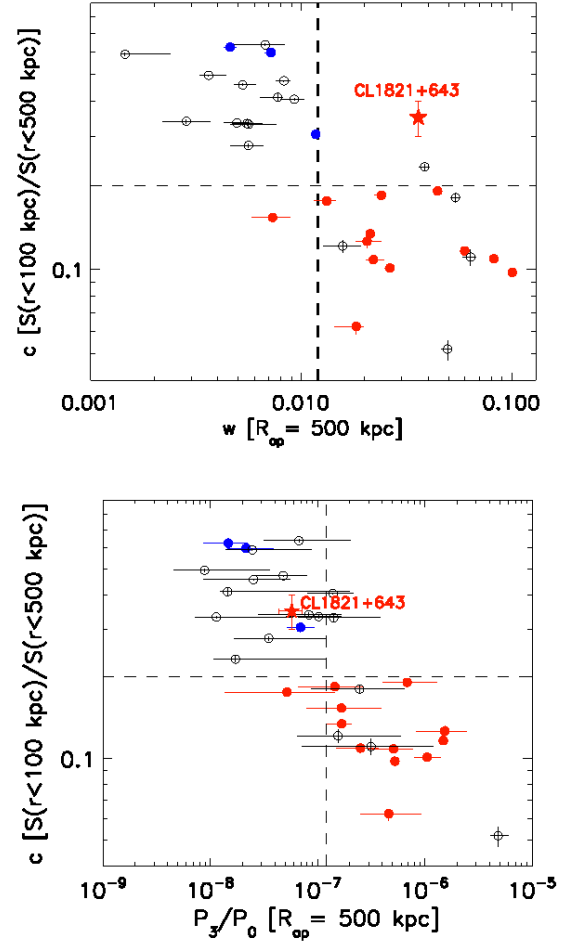}}
\end{picture}
\caption{The $c-w$ (top panel) and $c-P_3/P_0$ (bottom panel) diagrams. The red star marks the position of CL1821+643. Dashed lines are the median of the parameters, which define the radio-loud and radio-quiet quadrants, red dots mark giant radio halos, blue dots mark mini halos, black circles mark radio quiet clusters (points and lines are taken from from Cassano et al. 2010). }
\label{fig:mom}
\end{figure}

Since radio halos are always found in merging galaxy clusters, it is crucial to characterise the dynamical status of CL1821+643.
To this aim,  we compute three dynamical indicators, extensively used in the literature \citep[e.g.][]{Boehringer10}:  the power ratio $P_3/P_0$, the centroid shift, $w$, and the concentration parameter, $c$. Recently, \citet{Cassano10} have used these methods, finding that clusters with and without radio halos occupy different regions in the morphological diagrams (see Fig. \ref{fig:mom}), and confirming that radio halos are always associated to merging clusters.\\
\indent The {\it Chandra} image of the cluster CL1821+643 has been exposure, vignetting, and background corrected, and the
quasar emission has been subtracted.  $P_3/P_0$, $w$, and $c$ have been computed within an aperture radius $R_{ap}=500$ kpc. Because of the quasar and of the readout-strip, the centre of the cluster, used in the following analysis, is not uniquely defined.  We consider as centre the position of the quasar.\footnote{We repeated the calculations assuming different centres: (i)
the pixel that provides the minimum value of  the dipole $P_1$, (ii) the X-ray centroid, and (iii) the position where the X-ray emission is maximum. All these choices are affected by the quasar excision and, as such, less reliable. Nonetheless, the results we obtain differ by few \%.}  The errors and the photon bias have been estimated through Monte-Carlo simulations, following  the approach of \citet{Boehringer10}. \\
\indent The power ratio is a multipole decomposition of the two-dimensional projected mass distribution \citep{1995ApJ...452..522B}.  
We have determined the power ratio $P_3/P_0$, the lowest power ratio moment providing a clear substructure measure, obtaining $P_3/P_0=(5.8 \pm 1.4) \times 10^{-8}$. Since the momenta are computed excising the inner 0.05$\times R_{ap}$, the quasar does not affect the  $P_3/P_0$ computation.\\
\indent The centroid shift  method \citep{Maughan08} measures the standard deviation of the projected separation between the X-ray peak and the centroid in units of $R_{ap}$, computed within circles of increasing radius (from 0.05$\times R_{ap}$ to $R_{ap}$, in steps of 0.05$\times R_{ap}$). 
The centroid shift, $w$, has been computed minimising the $P_1$ dipole within circles of increasing radius (as done by \citealt{Boehringer10}), and computing the X-ray weighted centroid (following \citealt{Cassano10}). We obtain $w=(5.98 \pm 0.07) \times 10^{-2}$ and $w=(3.6  \pm 0.07) \times 10^{-2}$, respectively.\\
\indent The concentration parameter, $c$, measures the ratio between the X-ray surface brightness within 100 kpc over the surface brightness within 500 kpc. Because of the quasar and  the readout-strip, we can only derive limits for $c$, with and without the quasar subtraction. We obtain $ 0.3 <c<0.4$.\\
In Fig. \ref{fig:mom} we show the position of the cluster in the $P_3/P_0-c$, and $w-c$ diagrams, taken from \citet{Cassano10}. In the $P_3/P_0-c$ diagram, the cluster is among the radio-quiet ones. The $w$ parameter, instead, has a typical value for clusters with a giant radio halo, i.e. dynamically disturbed. Hence, in the $c$-$w$ and $w$-$P_3/P_0$ diagram (not shown here), the cluster is in a rather empty quadrant. 

\section{Discussion and conclusions}
\label{sec:discussion}
A giant-radio halo in a cool core cluster challenges our understanding of cool core clusters, major mergers, and radio emission. 
So far, giant radio halos have always been found in massive mergers, and never in cool core clusters, hence supporting the idea that
major mergers, necessary to power giant radio halos, also disrupt the cool core. 
In this framework, few peculiar cases have been found:  RXJ1347.5-1145, which shows both a cool core and merging signatures, and hosts a mini halo \citep{Gitti07}, and
A2142, where the cool core has been destroyed by the merger, but some remnants seem to be sloshing, and a giant radio halo has been detected  \citep{2013ApJ...779..189F}. CL1821+643 is the only case found so far  where a giant radio halo is detected in a cluster  that preserves the cool core.  \\
\indent So far, giant radio halos and mini halos have been considered as two separate classes of sources. 
CL1821+643 could be witnessing the transition between these two stages: turbulent motions generated during mergers could switch off the mini halo and transport relativistic electrons on larger scales \citep{BJ14}. 
In this case, one  should assume that a merger is taking place here, and that it has not disrupted the cool core.\\
\indent The X-ray observations indicate that the cluster is not undergoing a major merger similar to those detected in other clusters with radio halos. However, it could be undergoing a minor or off-axis merger, which may be responsible for the high value of the centroid shift $w$ and the possible sloshing structure (Russell et al. 2010).
As none of the morphological indicators is sensitive to mergers along the line of sight, spectroscopic observations are needed to finally assess the dynamical status of CL1821+643. The analysis of the X-ray substructure  highlights the peculiarities of CL1821+643, even without the radio information. Indeed,  
 it shows features typical of systems in dynamical equilibrium (cool core, temperature and entropy drop, high value of the concentration parameter $c$) as well as indication of large-scale perturbations (high value of the centroid shift $w$ and sloshing).\\
\indent Another possibility is that the radio emission is not a radio halo but a radio relic seen in projection. Radio relics could be originated by minor off-axis mergers that preserve the cool core, like in Abell 1664 \citep{2001A&A...376..803G}. However, the presence of a radio relic seen in projection onto the cluster centre is unlikely \citep{Vazza12}
and neither the morphology of the emission nor the NW elongation  - roughly following the X-ray emission - would favour this interpretation.\\
 Finally, the extended emission could result from the past activity of a FRII radio source, now turned into the FRI
 associated with the quasar, hence not requiring a merger to accelerate the particles.
 We regard this possibility as less plausible because of the relatively low $\alpha$, and the morphology of the radio emission. However, spectral index maps would be needed to investigate this scenario.\\
\indent To conclude, our results can be summarised as follows:\\
 $\bullet$ We have analysed the radio emission of the cluster CL1821+643 - also known as PSZ1\,G094.00+27.41 - finding for the first time a strong cool core cluster that hosts a giant radio halo, having a largest linear scale of $\sim$ 1.1 Mpc.\\
$\bullet$ We have constrained the spectral index between 323 MHz and 1.665 GHz to be $1.04 \leq \alpha \leq 1.1$, meaning that the halo is in an early stage of its radio-loud phase. \\
$\bullet$ The radio power at 1.4 GHz  is at least  a factor $\sim$3 below the $P_{\rm 1.4 \,GHz}- L_{\rm [0.1-2.4] \, keV}$ correlation, while it sits on the $\rm{SZ}-P_{\rm 1.4 \,GHz}$ correlation.\\
$\bullet$ An analysis of the cluster, based on the X-ray $P_3/P_0$ and  $c$  morphological estimators would classify the cluster among the radio-quiet systems. However, the high value of $w$ indicates that some merger has happened here.\\
 \indent If the radio emission is powered by merger-induced turbulence, we have to assume that  the cluster is undergoing a merger which has not disrupted the cool core, but has injected a sufficient amount of energy to power the radio emission.
  In re-acceleration scenarios, this suggests that a higher fraction of the gravitational energy, released into the ICM during mergers, might be channeled into the (re)acceleration of particles, or that a seed population of energetic particles is already  present.
 
\section*{Acknowledgments}
We thank F. de Gasperin and F. Vazza for useful discussions. A.B and M.B  acknowledge support by the research group FOR 1254 funded by the Deutsche Forschungsgemeinschaft. H.R. acknowledges support from ERC Advanced Grant Feedback. R.J.W. is supported by NASA through Einstein Postdoctoral grant PF2-130104 (Chandra X-ray Center, SAO for NASA, contract NAS8-03060). We thank the staff of the GMRT. NRAO is a facility of the National Science Foundation operated under Associated Universities Inc.

\bibliographystyle{mn2e}
\bibliography{master}

\end{document}